\documentclass[twocolumn]{aastex63}

\received{29-May-2021}
\revised{14-Jul-2021}
\accepted{24-Jul-2021}
\shorttitle{Searching for low-redshift faint galaxies with MMT/Hectospec}
\shortauthors{Cheng et al.}
\graphicspath{{./}{figures/}}

\begin{document}

\title{Searching for low-redshift faint galaxies with MMT/Hectospec}

\correspondingauthor{Cheng Cheng}
\email{chengcheng@nao.cas.cn}

\author{Cheng Cheng}
\affiliation{Chinese Academy of Sciences South America Center for Astronomy, National Astronomical Observatories, CAS, Beijing 100101, China}
\affiliation{CAS Key Laboratory of Optical Astronomy, National Astronomical Observatories, Chinese Academy of Sciences, Beijing 100101, China}

\author{Jia-Sheng Huang}
\affiliation{Chinese Academy of Sciences South America Center for Astronomy, National Astronomical Observatories, CAS, Beijing 100101, China}
\author{Christopher N. A. Willmer}
\affiliation{Steward Observatory, University of Arizona, 933 N. Cherry Avenue, Tucson, AZ 85721, USA}
\author{Hong-Xin Zhang}
\affiliation{CAS Key Laboratory for Research in Galaxies and Cosmology, Department of Astronomy, University of Science and Technology of China, Hefei, Anhui 230026, China}
\affiliation{School of Astronomy and Space Science, University of Science and Technology of China, Hefei 230026, China}

\author{Matthew L. N. Ashby}
\affiliation{Center for Astrophysics $|$ Harvard and Smithsonian, 60 Garden St, Cambridge, MA 02138, USA}

\author{Hai Xu}
\affiliation{Chinese Academy of Sciences South America Center for Astronomy, National Astronomical Observatories, CAS, Beijing 100101, China}

\author{Marcin Sawicki}
\affiliation{Institute for Computational Astrophysics and Department of Astronomy and Physics, Saint Mary's University, 923 Robie Street, Halifax, Nova Scotia, B3H 3C3, Canada}

\author{Stephane Arnouts}
\affiliation{Aix Marseille Universit\'e, CNRS, Laboratoire d’Astrophysique de Marseille, UMR 7326, F-13388 Marseille, France}

\author{Stephen Gwyn}
\affiliation{Canadian Astronomy Data Centre, NRC-Herzberg, 5071 West Saanich Road, Victoria, British Columbia V9E 2E7, Canada}

\author{Guillaume Desprez}
\affiliation{Department of Astronomy, University of Geneva, ch. d'\'Ecogia 16, CH-1290 Versoix, Switzerland}

\author{Jean Coupon}
\affiliation{Department of Astronomy, University of Geneva, ch. d'\'Ecogia 16, CH-1290 Versoix, Switzerland}

\author{Anneya Golob}
\affiliation{Institute for Computational Astrophysics and Department of Astronomy and Physics, Saint Mary's University, 923 Robie Street, Halifax, Nova Scotia, B3H 3C3, Canada}

\author{Piaoran Liang}
\affiliation{Chinese Academy of Sciences South America Center for Astronomy, National Astronomical Observatories, CAS, Beijing 100101, China}

\author{Tianwen Cao}
\affiliation{Chinese Academy of Sciences South America Center for Astronomy, National Astronomical Observatories, CAS, Beijing 100101, China}

\author{Yaru Shi}
\affiliation{Chinese Academy of Sciences South America Center for Astronomy, National Astronomical Observatories, CAS, Beijing 100101, China}

\author{Gaoxiang Jin}
\affiliation{Chinese Academy of Sciences South America Center for Astronomy, National Astronomical Observatories, CAS, Beijing 100101, China}

\author{Chuan He}
\affiliation{Chinese Academy of Sciences South America Center for Astronomy, National Astronomical Observatories, CAS, Beijing 100101, China}

\author{Shumei Wu}
\affiliation{Chinese Academy of Sciences South America Center for Astronomy, National Astronomical Observatories, CAS, Beijing 100101, China}

\author{Zijian Li}
\affiliation{Chinese Academy of Sciences South America Center for Astronomy, National Astronomical Observatories, CAS, Beijing 100101, China}

\author{Y. Sophia Dai}
\affiliation{Chinese Academy of Sciences South America Center for Astronomy, National Astronomical Observatories, CAS, Beijing 100101, China}

\author{C. Kevin Xu}
\affiliation{Chinese Academy of Sciences South America Center for Astronomy, National Astronomical Observatories, CAS, Beijing 100101, China}

\author{Xu Shao}
\affiliation{Chinese Academy of Sciences South America Center for Astronomy, National Astronomical Observatories, CAS, Beijing 100101, China}

\author{Marat Musin}
\affiliation{Chinese Academy of Sciences South America Center for Astronomy, National Astronomical Observatories, CAS, Beijing 100101, China}


%
%
%
%




\begin{abstract}
We present redshifts for 2753 low-redshift galaxies between $0.03 \lesssim z_{\rm spec}\lesssim0.5$ with  18 $\leq$ $r$ $\leq$ 22 obtained with Hectospec at the Multi-Mirror Telescope (MMT). The observations targeted the XMM-LSS, ELAIS-N1 and DEEP2-3 fields, each of which covers $\sim$ 1 deg$^2$.  These fields are also part of the recently completed CFHT Large Area U-band Deep Survey (CLAUDS) and on-going Hyper Suprime-Cam deep fields surveys. The efficiency of our technique for selecting low-redshift galaxies is confirmed by the redshift distribution of our sources. In addition to redshifts, these high S/N spectra are used to measure ages, metallicities, and nuclear activity levels.  In combination with the photometric catalogue in $u$, $g$, $r$, $i$, $z$, $y$ down to 27 AB mag, we are able to study the galaxy population down to stellar masses of $\sim$ 10$^8 M_\odot$ . This paper presents the observational strategy, the reduction procedure and properties of the galaxy sample\footnote{The catalog can be accessed through the webpage of this survey: {\url{http://mips.as.arizona.edu/~cnaw/Faint_Low_z/}}.}.
\end{abstract}




\section{introduction}
Galaxy populations are bimodally distributed in colour, morphology, metallicity and so on, which indicates a divergence in the galaxy evolution path \citep{Takamiya1995, Baldry2004}. The observational evidence indicates that galaxies evolve from the blue cloud to the red sequence \citep[e.g., ][]{Faber2007}. For the bright galaxies, the population of blue and red can be defined well, however, whether the bimodality relation can be extended to the faint end is still unclear. 

Current surveys such as Hyper Suprime-Cam (HSC) Subaru Strategic Program \citep[HSC-SSP,][]{Aihara2018a,Aihara2018b,Aihara2019} and CFHT Large Area U-band Deep Survey \citep[CLAUDS,][]{Sawicki2019} 
provide catalogs that simultaneously cover large areas and reach faint magnitude limits. The finished CLAUDS project on the field of XMM-Newton Large-Scale Structure \citep[XMM-LSS,][]{2007MNRAS.382..279P},
The European Large Area ISO Survey N1 area \citep[ELAIS-N1,][]{2004MNRAS.351.1290R},
Deep Extragalactic Evolutionary Probe 2, Field 3 \citep[DEEP2-F3,][]{2013ApJS..208....5N},
and Cosmic Evolution Survey \citep[COSMOS,][]{2007ApJS..172....1S}, produces $u$ band catalogues with about 20 deg$^2$ and deep to 27 AB mag, which is about 2 mags deeper than previous large area surveys like Canada-France-Hawaii Telescope Legacy Survey \citep[CFHTLS,][]{2012SPIE.8448E..0MC}. The CLAUDS area is also covered by the HSC deep survey project with $u$, $g$, $r$, $i$, $z$, $y$ bands at a depth of 27 AB mag. The matched CLAUDS+HSC catalogue provides us with an unprecedented chance to study the faint galaxies statistically. 

Almost all of our current understanding of the galaxy formation and evolution, such as the bimodality of the galaxy colour, the luminosity or stellar mass functions require accurate spectroscopic redshift (spec-$z$) measurements. Thus the spectroscopic redshift surveys are the key to understand the formation and evolution of galaxies. The advantage of the current wide field deep survey is that we will have both a large sample with bright luminosities at high redshift and a large faint galaxy sample at lower redshift.

Obtaining statistically significant samples of low luminosity galaxies is rather difficult. Because of their faintness, the number of spectroscopically observed dwarf galaxies is very small even in large surveys like Sloan Digital Sky Survey \citep[SDSS,][]{2000AJ....120.1579Y, 2002AJ....124.1810S, 2006AJ....131.2332G} 
or Galaxy and Mass Assembly \citep[GAMA,][]{2011MNRAS.413..971D}. This is particularly true for galaxies in the red sequence. Deep field imaging surveys \citep[such as ][]{Ilbert2010, Huang2013} suggest that there is a deficit of red galaxies at the low mass end. This bias becomes more important in optical surveys with bright limiting magnitudes. To have a complete census of the low mass galaxy population, we need to have redshifts of these faint galaxies.

Constructing a large faint-galaxy sample requires a redshift survey that combines large area coverage with a deep magnitude limit. However, one dilemma of the spec-$z$ survey is the difficulty of getting a large areal coverage and reaching very faint simultaneously. SDSS as the most extensive spec-$z$ survey has covered 3600 deg$^2$ and reached the spec-$z$ complete depth at $r \sim$ 17.7 mag. The GAMA spec-$z$ survey reaches a depth of $r \sim$ 19.8 AB mag, and is complete in stellar masses to $\sim 10^9M_\odot$. Nevertheless, the spectra of intrinsically very faint galaxies are often too noisy to be identified with high confidence or are biased towards containing predominantly emission-line galaxies, so surveys such as GAMA could be biased towards over-representing emission-line galaxies, and miss the low-redshift faint red galaxy population at $10^9M_\odot$.

On the other hand, photometric redshifts for faint blue galaxies at lower redshifts suffer very large uncertainties due to their faintness and the flat Spectral Energy Distribution (SED) shape \citep{Dahlen2013}. Moreover, to obtain a reliable photometry redshift (phot-$z$) estimation, we also need a spec-$z$ sample with a similar magnitude limit to estimate the accuracy of the phot-$z$ results. Artificial Intelligence (AI) shows great promise in obtaining the phot-$z$s with unprecedented accuracy and efficiency \citep{2018PASJ...70S...9T}. However, AI also requires a large training sample with spec-$z$ containing a variety of galaxy populations and brightnesses. Therefore, although we can estimate the redshift from the photometric catalogues, we still need spec-$z$ to calibrate or training the phot-$z$ results.

Nevertheless, because of constraints imposed by the efficiency of spectroscopic observations, spec-$z$ survey projects are biased in the sense of containing mainly intrinsically bright objects. For the four CLAUDS fields, only the COSMOS field has the spec-$z$ survey projects \citep[zCOSMOS,][]{2007ApJS..172...70L, 2009ApJS..184..218L} for both low and high-redshift galaxies with a depth about $i \sim 22.5$ AB mag \citep[also see][]{2018ApJS..234...21D}, and the XMM-LSS field has been covered by the VIMOS VLT deep survey \citep[VVDS,][]{2005A&A...439..845L} to about $i \sim $24 with low spectral resolution \citep{2013A&A...559A..14L}. The DEEP2-3 field is covered by the DEEP2 survey \cite{2013ApJS..208....5N}, which only focuses on the $z_{\rm spec}>0.5$ galaxies. The ELAIS-N1 field was selected from the SIRTF Wide-Area Infrared Extragalactic Survey \citep[SWIRE,][]{2003PASP..115..897L}, and is mainly covered by SDSS dr13 for the bright targets \citep{2013MNRAS.428.1958R, 2013ApJS..208...24L, 2021arXiv210505659S}. Thus, the CLAUDS fields still lack dense spectroscopic coverage for fainter targets.

To investigate the galaxy properties for the low-redshift\footnote{
In this paper, we refer the galaxies at $0.03<z_{\rm spec}<0.5$ as low-redshift galaxies, and the galaxies at $z_{\rm spec}<0.03$ as local galaxies} low mass galaxies (e.g., $M_* \sim 10^{8.5}M_{\odot}$), we need a complete sample down to $r\simeq 22$ AB mag\footnote{Galaxy stellar mass about $M_* \sim 10^{8.5}M_{\odot}$ corresponding to about $M_r \simeq -17$ AB mag, which is about $r$ band 22 AB mag at redshift 0.15 \citep[][]{Mahajan2018}.}. The existing spectroscopic redshift surveys, such as SDSS, zCOSMOS or VIMOS Ultra-Deep Survey \citep[UVDS,][]{2015A&A...576A..79L}, are either shallow with large areas, or deep enough within a small area. Previous low mass galaxy spectroscopic redshift survey such as Metal Abundances across Cosmic Time (MACT) Survey \citep{2016ApJS..226....5L, 2016ApJ...828...67L, 2021MNRAS.501.2231S} only focuses on the emission line galaxies, which would bias to the young, star-forming galaxies. The ongoing spec-$z$ survey projects such as Dark Energy Spectroscopic Instrument \citep[DESI,][]{2016arXiv161100036D}, Subaru Prime Focus Spectrograph \citep[PFS,][]{2016SPIE.9908E..1MT} and Hobby-Eberly Telescope Dark Energy Experiment \citep[HETDEX,][]{2008ASPC..399..115H} plan to cover large areas with much deeper survey depth, but these survey projects are very time consuming or are only initiating. Therefore, we propose a new redshift survey for the low-redshift faint galaxies based on the CLAUDS+HSC catalogue with MMT/Hectospec.

MMT/Hectospec \citep{2005PASP..117.1411F} as a multi-fibre spectrograph that can obtain 300 fibre spectra simultaneously within 1 degree diameter, and thus very efficient for redshift surveys. Moreover, for the low-redshift galaxies, the moderate resolution spectrum (resolution power about 2700) covering from 3600\AA\ to 8900\AA\ will reveal most of the optical spectrum lines such as [OII], Ca II,  H$\beta$, [OIII], Mgb, Na D, H$\alpha$, [NII], [SII], which are not only helpful in identifying the redshift, but also necessary to characterize the physical properties such as star formation rate, metallicity, galactic activity, etc.

To study the galaxy population especially the low-redshift faint field galaxies, we performed a redshift survey in the HSC-survey deep fields by MMT/Hectospec. Previous low-redshift redshift surveys such as Satellites Around Galactic Analogs \citep[SAGA,][]{2017ApJ...847....4G, 2021ApJ...907...85M} and Southern Stellar Stream Spectroscopic Survey \citep[$S^5$,][]{2019MNRAS.490.3508L} focused on the satellite galaxies and stellar streams at $z_{\rm spec} \lesssim 0.02$, the properties of which would be affected by the environment. There are very few local bright galaxies ($z_{\rm spec} < 0.02$) in the HSC deep fields, because the deep fields are selected to study the galaxies at high-$z$, thus we can expect the redshift survey in HSC deep fields can provide us the low-redshift faint field galaxies at $z_{\rm spec} \gtrsim 0.02$. The HSC+CLAUDS provide us with a photometric catalog reaching to 27 AB mag, which is faint enough for us to select low-redshift faint galaxy candidates, and the MMT 6.5 meter mirror can push the redshift survey depth to about $r$ = 22 AB mag, which is roughly corresponds to stellar masses of $10^{8.5}M_\odot$ at redshift 0.15. This sample reaches about 4 magnitudes deeper than SDSS, and two mags deeper than GAMA survey, which will be critically important to understand the faint galaxy properties and the calibration of the phot-$z$ to spec-$z$ to the faint end.

\begin{figure*}[ht!]
\centering
\includegraphics[width=0.95\textwidth]{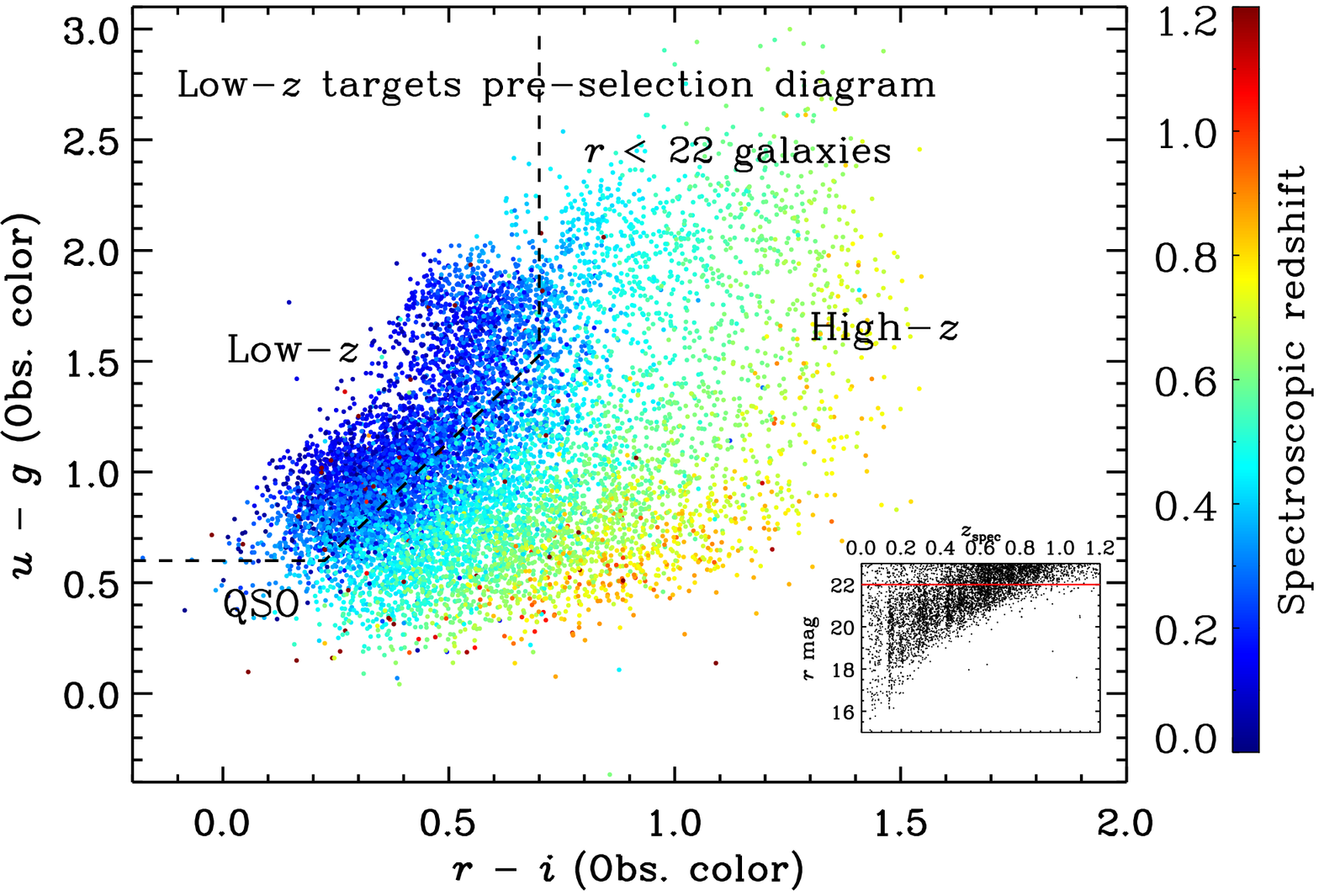}
\caption{The low-$z$ targets pre-selection method. We show data from XMM-LSS field as an example. The colour code
represents redshifts from prior surveys including SDSS, VIMOS VLT deep survey \citep[VVDS,][]{2005A&A...439..845L}, the VIMOS Public Extragalactic Survey \citep[VIPES,][]{2018A&A...619A..17M}, PRIsm MUlti-object Survey\citep[PRIMUS,][]{2011ApJ...741....8C} and eBoss\citep[][]{2016AJ....151...44D}. From the main plot, we can see a clear trend that redshifts change with colour. We remove the blue QSOs in the lower-left corner by setting $u-g > 0.6$. Our selection criteria ($  u-g > 0.6; r - i < 0.7; u - g > 1.95 \times (r-i) + 0.16 $) are shown by the dashed lines, which select most of the galaxies at $z_{\rm spec}<0.35$.The inset in the lower right shows the distribution of the $r$ band magnitude and redshift. The red line shows the $r=22$ limit, which enables us to remove the contamination from the high-$z$ galaxies.} \label{ugri_selection}
\end{figure*}

\begin{figure*}
    \centering
    \includegraphics[width=0.99\textwidth]{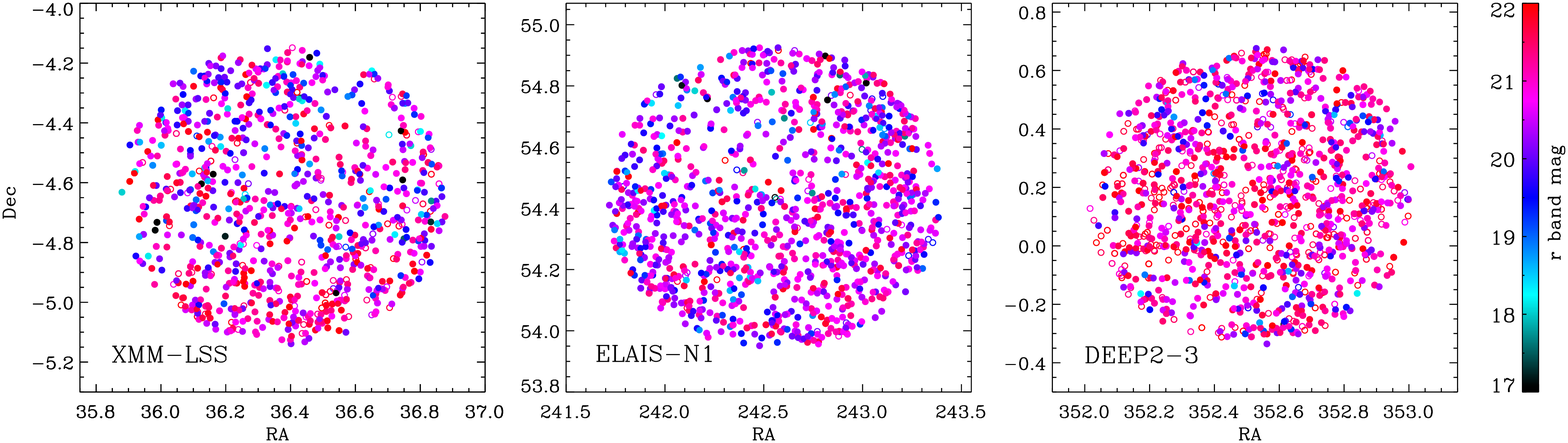}
    \caption{Survey region in XMMLSS, ELAIS-N1 and DEEP2-3. We colour the target position by the $r$ band magnitude. The filled circles show the targets with reliable redshifts ($qz\ge3$) while the open circles represent observed galaxies that did not provide a reliable redshift.}
    \label{region}
\end{figure*}

In this paper, we describe our survey strategy, redshift catalogues and preliminary results of this survey. We arrange this paper as follows: Section 2 describes the target selection procedure, the MMT/Hectospec observations and the  data reduction method. In Section 3, we show the typical spectrum examples as well as the first look results. In close with a summary in Section 4.

\section{Observations and Data Reduction}

\subsection{Target selection}

\begin{figure}[ht!]
\centering
\includegraphics[width=0.43\textwidth]{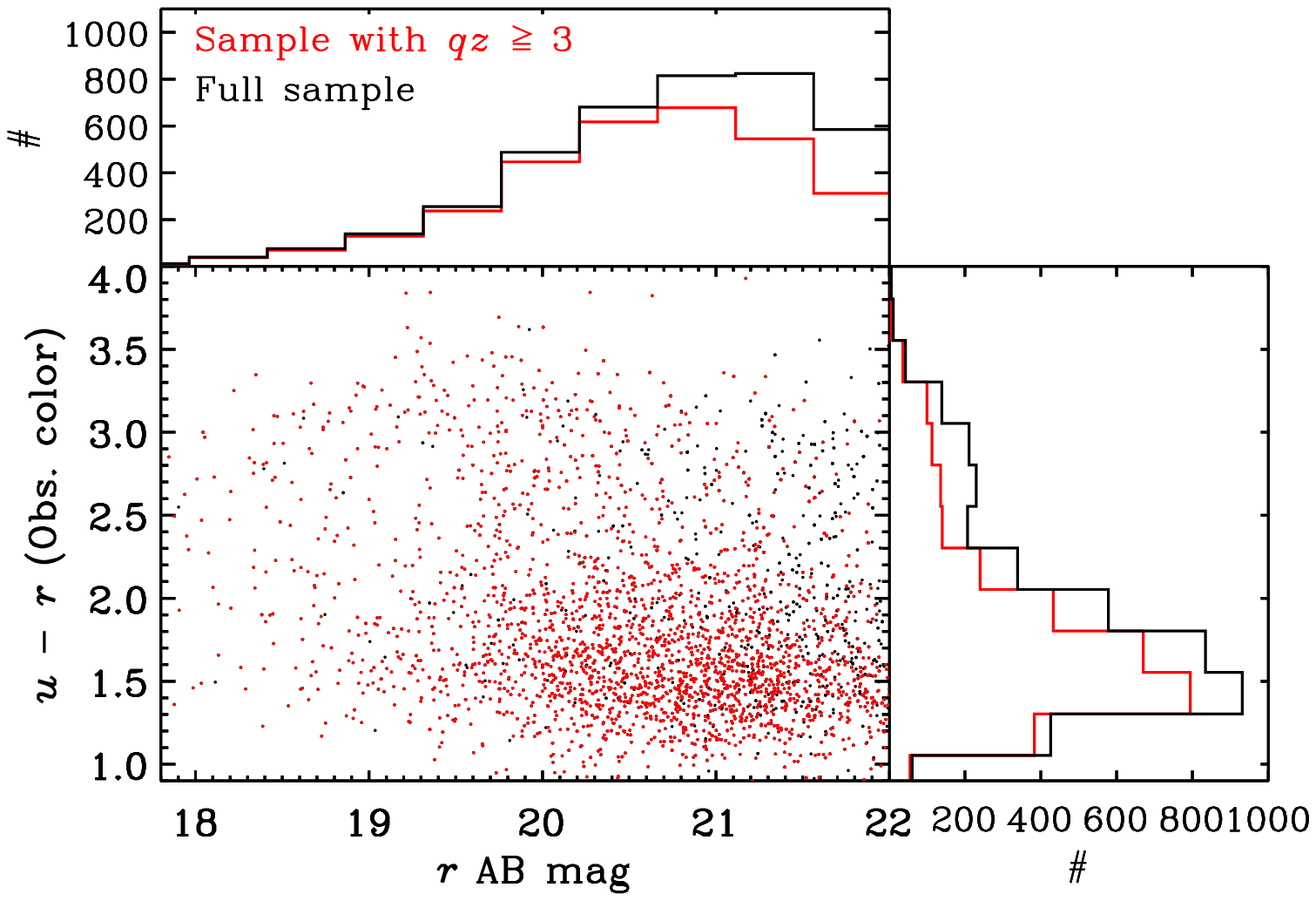}
\includegraphics[width=0.47\textwidth]{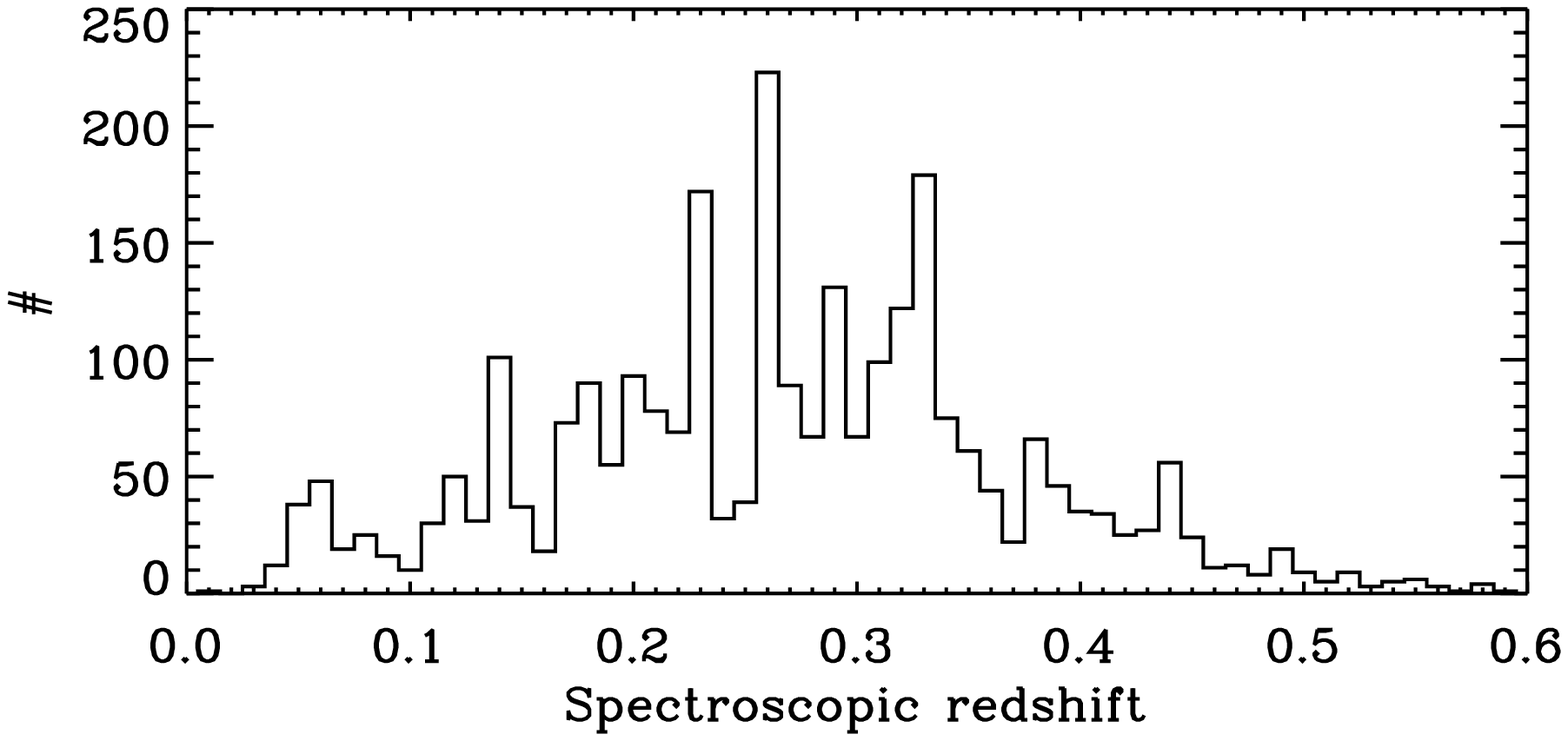}
\caption{Upper panel: observed color magnitude diagram of our full sample observed by MMT/Hectospec. We highlight the galaxies with redshift quality $qz \ge 3$ in red color. The spec-$z$ identification rate is the ratio of the red and black histograms, which is about 50\% for the galaxies about $r$ band 22 AB magnitude, or $u-r \gtrsim 2.5$.
Lower panel: The redshift distribution of our final results.
} \label{specz_hist}
\end{figure}
\begin{figure*}[ht!]
\centering
\includegraphics[width=0.95\textwidth]{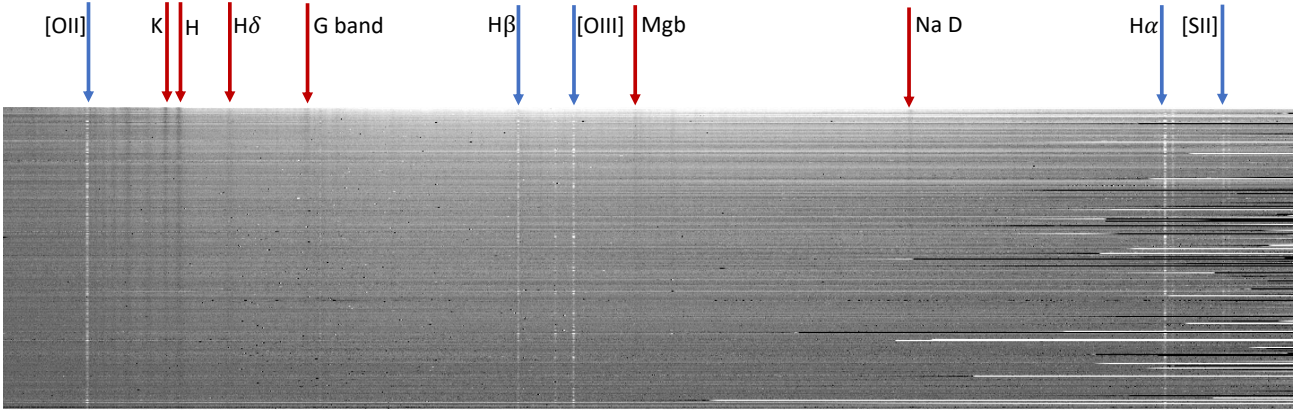}
\caption{All one-dimensional galaxy MMT/Hectospec spectra for which reliable redshifts could be estimated, displayed in de-redshifted two-dimensional format with major features noted. 
}\label{spec-image}
\end{figure*}

\begin{figure}[ht!]
\centering
\includegraphics[width=0.44\textwidth]{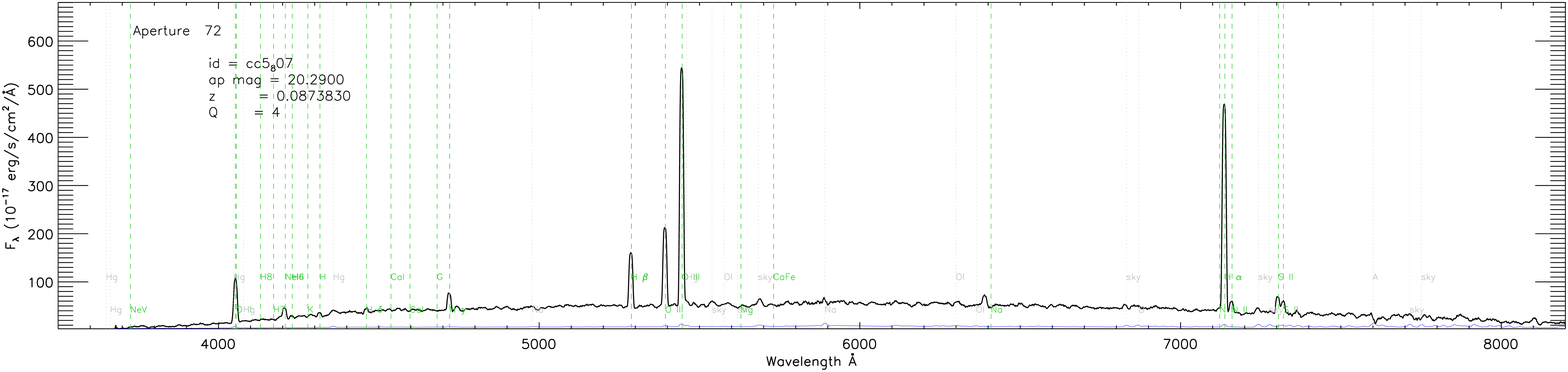}
\includegraphics[width=0.44\textwidth]{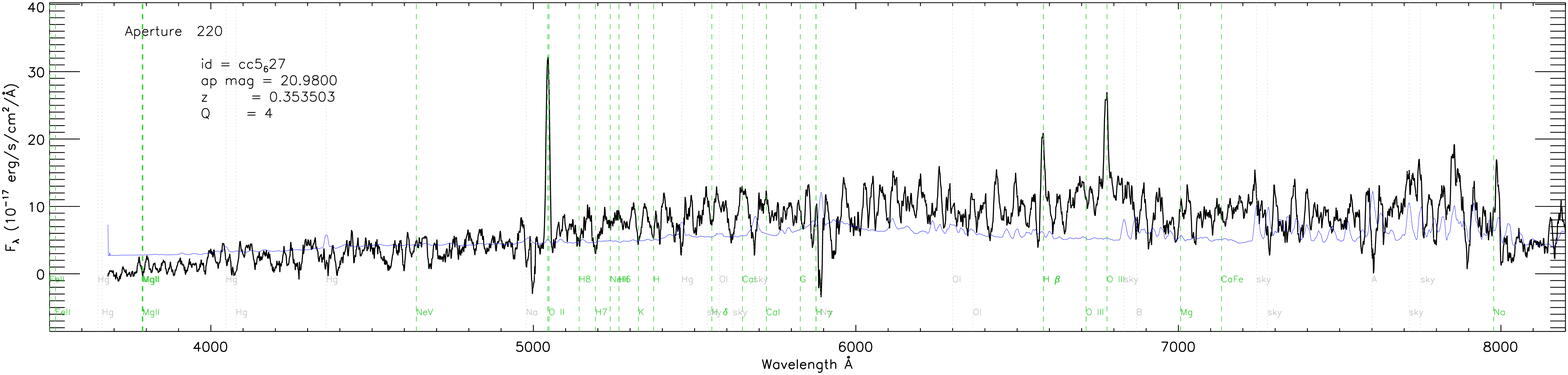}
\includegraphics[width=0.44\textwidth]{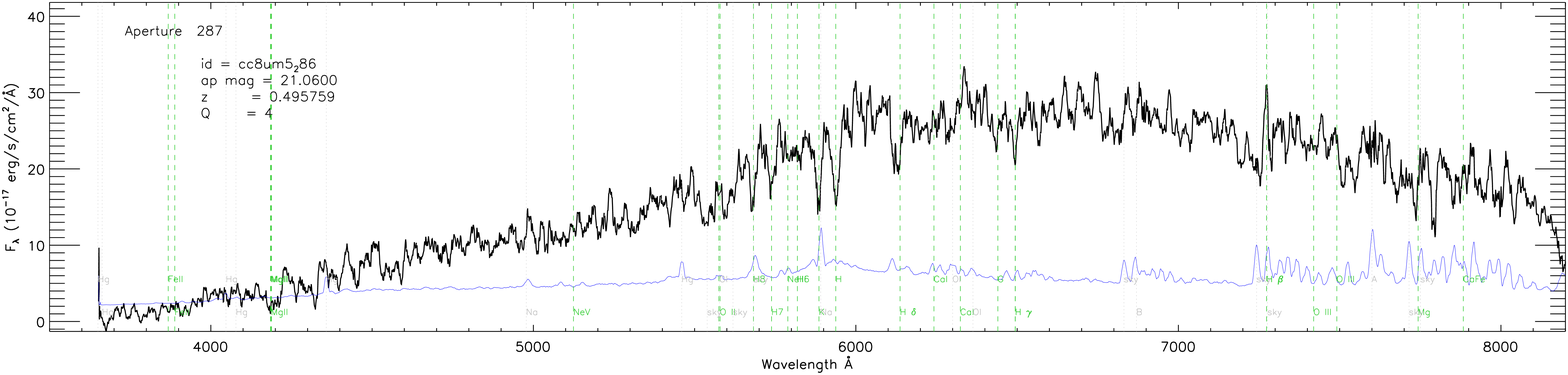}
\includegraphics[width=0.44\textwidth]{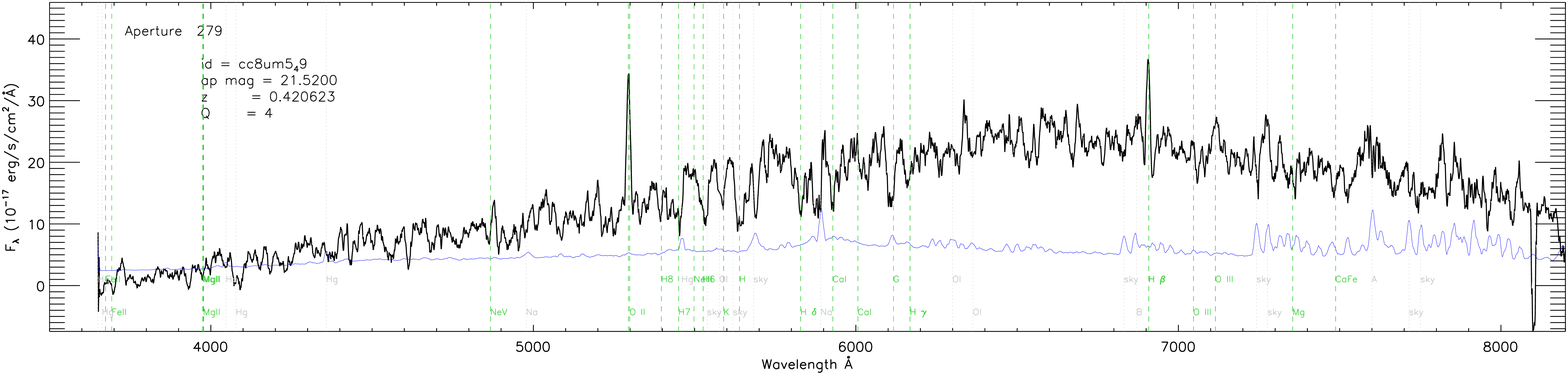}
\includegraphics[width=0.44\textwidth]{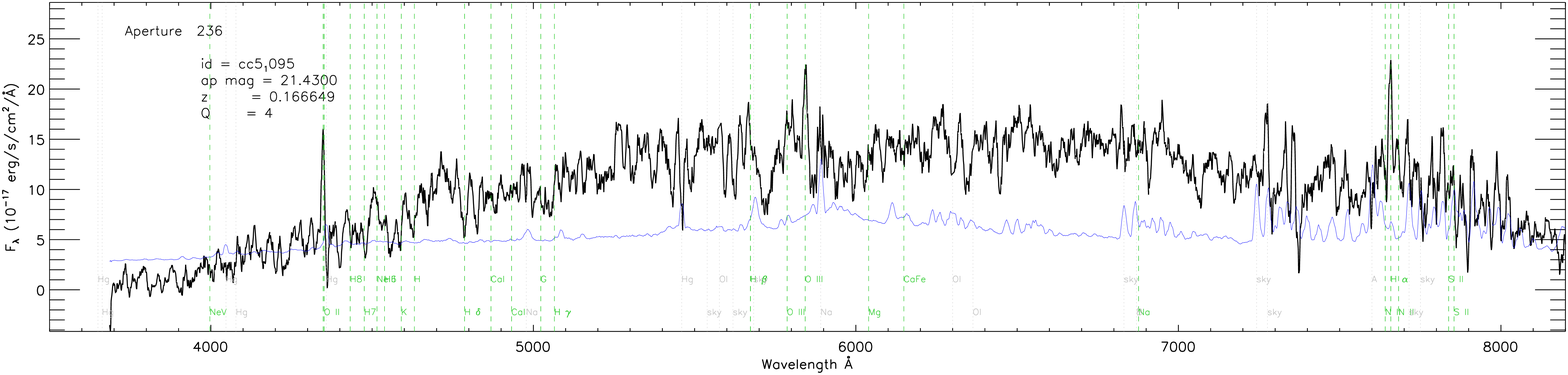}
\includegraphics[width=0.44\textwidth]{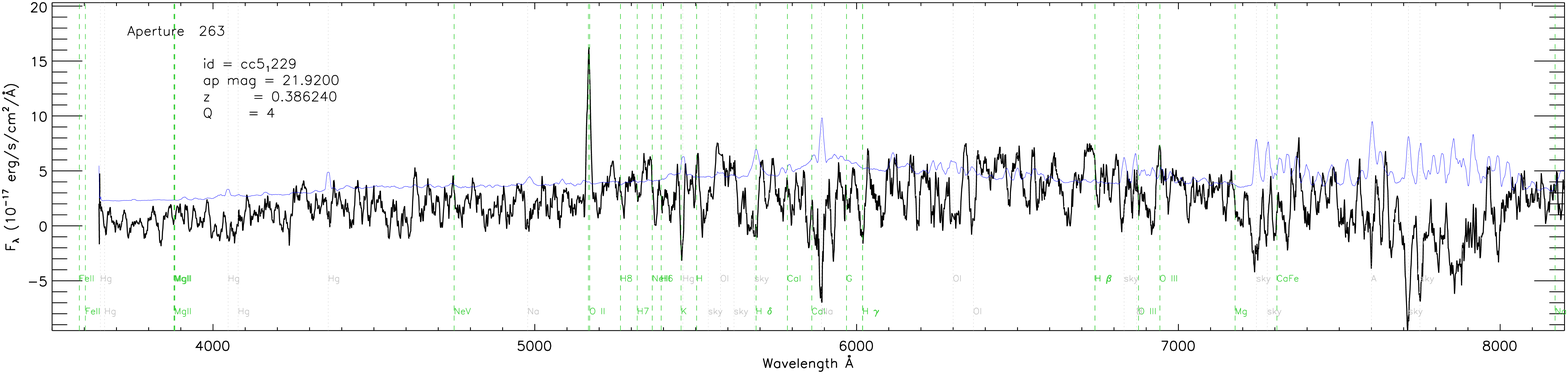}
\caption{Examples of MMT/Hectospec spectra. The $r$ band magnitudes increase from top to bottom. The green dashed lines show the emission or absorption line positions. The black lines are the galaxy spectra, while the blue lines are the inverse sqrt(variance) spectra. We denote the target name, $r$ band aperture magnitude within 1", identified redshift and the redshift quality ($qz$) in each panel. Redshifts for galaxies brighter than $r = 21$ can be identified clearly. For the galaxies fainter than $r = 21$, redshift of galaxies with bright emission lines can be identified. The bottom panel shows no clear continuum due to the faintness, while the emission lines still help us to measure the redshift.
} \label{spec-example}
\end{figure}

The low-redshift galaxy population with stellar mass $\sim 10^8M_\odot$ should have $r$ $\sim$ 22 at $z_{\rm spec}\simeq 0.15$, thus we select the galaxies with $ r < 22$. The four CLAUDS fields are covered by SDSS, thus the galaxies with $r < 17.7$ have complete spec-$z$. So we limit our targets to $18< r<22$. When the observations were initiated, the CLAUDS survey was essentially complete, with full coverage in XMM-LSS and ELAIS-N1, while HSC data were not yet fully released. Since the $r$ band 22 magnitude is still above the limit magnitude of the previous survey projects, we selected our targets from the CFHTLS catalogue \citep{2012AJ....143...38G} for the CLAUDS-XMMLSS field, ELAIS-N1 catalogue \citep{2011MNRAS.416..927G} for the CLAUDS-ELAIS-N1 field and the Stripe-82 catalogue\citep{2014ApJS..213...12J} for the CLAUDS-DEEP23 field. The minor differences between the photometric passband definitions will not affect out target selection.

To target specifically galaxies at lower redshifts, we adopt a colour-colour cut, similar to the method used in the DEEP2 project \citep{2013ApJS..208....5N} to pre-select our targets. Fig. \ref{ugri_selection} illustrates the selection methods. Galaxies with $z_{\rm spec}<0.35$ are located in the upper left of the $u-g$, $r-i$, colour-colour diagram, and are separated into two groups, indicating a blue and red population in colours. The inset in Fig. \ref{ugri_selection} show the $r$ band magnitude and redshift distribution. So excluding galaxies fainter than $r=22$\,AB mag removes most higher-redshift galaxies. We select our sample galaxies by the criteria: $  u-g > 0.6; r - i < 0.7; u - g > 1.95 \times (r-i) + 0.16 $. The CLAUDS catalogue is deep enough for us to select the $u-g$ red galaxies, which are candidate low mass red galaxies. We select the upper left corner of the dashed line selection criterion in Fig. \ref{ugri_selection}. We also remove the foreground stars by the CLASS\_STAR flag, and the galaxies with previous spec-$z$ measurements. The survey was designed to observe galaxies brighter than $r$=22.5, which is the detection limit for galaxies in 1 hour integrations. Therefore, the MMT/Hectospec is exactly the instrument suitable for this project.

We were allocated 6.5 MMT/Hectospec nights  (= 4.5 dark nights + 2 grey nights) from the Telescope Access Program (TAP) to observe the XMM-LSS, ELAIS-N1 and DEEP2-3 fields (TAP2016B, ID:21; TAP2017A, ID: 27, TAP2017B, ID: 33. PI: Cheng Cheng). Hectospec is an optical fiber-fed spectrograph that can obtain 300 spectra at the same time in 1 $\rm deg^2$ Field-of-View for one configuration. Fig. \ref{region} shows the location of our targets in the three CLAUDS fields. We chose the 
pointing positions that were covered by previous spec-$z$ surveys to increase the survey completeness. Since the density of the $r \lesssim 22$ targets at low-$z$ is about 2000 per degree \citep{2004AJ....127.3121W}, we chose only one pointing per field with several configurations. The observations are run in queue mode and the user provides a previously calculated configuration file containing for each object the R.A. and Dec. positions, magnitudes and rank, using the {\tt xfitfibs}\footnote{\url{https://www.cfa.harvard.edu/mmti/hectospec/xfitfibs/}} code developed by the Hectospec Instrument team \citep{1998SPIE.3355..324R}. For each run, we assigned at least 40 fibres for the sky, 10 fibres for F-star flux calibrators and 3 fibres for the Guide stars. {\tt xfitfibs} will assign the fibre by the rank and class of each target. To avoid fibre entanglement, some of the fibres may not be assigned to an object. So we also add some $8\mu$m detected sources from the SWIRE survey with low rank to make full use of the available fibres. {\tt xfitfibs} calculates several configurations in a single field simultaneously in order to sample as many objects as possible in a given rank (i.e., priority). As each object is assigned to a fibre, it is ``removed" from the catalogue. To obtain longer exposures (i.e., over more than one configuration), some faint $u-g$ red sources were included more than once in the source catalogue used by {\tt xfitfibs}. But we found this method does not provide us with better S/N because of the low flux calibration accuracy and different weather at each night.

The observations were obtained in 2016 Oct. for XMMLSS and DEEP2-3 fields (2.5 dark nights), 2017 April for ELAIS-N1 (2 dark nights) and 2017 Oct. for XMMLSS and DEEP2-3 fields (2 grey nights). Due to the weather and instrument malfunction, we only succeeded in obtaining raw data for 20 $\times$ 1.5 hours exposure in total, which is about 4 nights data.

\begin{figure}[ht!]
\centering
\includegraphics[width=0.48\textwidth]{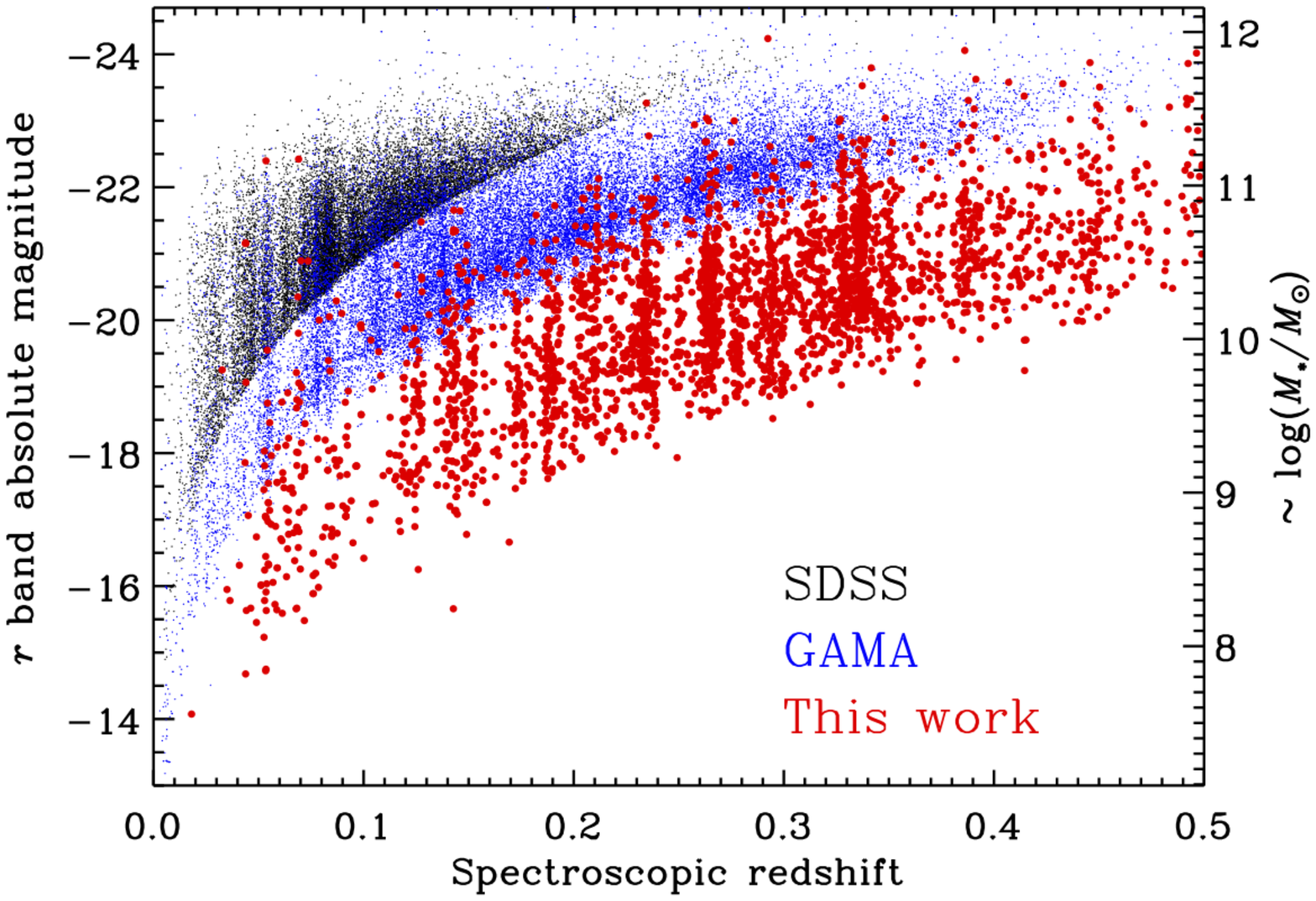}
\caption{Redshift and absolute mag plot of this spectroscopic redshift survey compared to the SDSS and GAMA surveys. At the same redshift, our survey is about 2 mag deeper than the previous surveys. For the galaxies at about $M_r\simeq -18$, the SDSS and GAMA sample only have the galaxies at $z_{\rm spec}<0.05$, while our results have the galaxies at $z_{\rm spec}<0.2$, where peculiar motions can be neglected. The right side of y-axis shows the stellar mass based on the mass-light relation from the low redshift red galaxies ($u-r>2$) in the COSMOS multi-wavelength catalog \citep{2016ApJS..224...24L}.}\label{abs_mag}
\end{figure}

\subsection{MMT/Hectospec data reduction}
The data are reduced by the HSRED2 pipeline \footnote{\url{http://mmto.org/~rcool/hsred/index.html}}, which will automatically perform the wavelength calibration, flux calibration, spectral extraction, etc. from the raw data, and produce the calibrated spectra. The flux is calibrated by the 10 F-star simultaneously obtained for each run. The final redshifts are identified by the {\tt spec2d} pipeline \footnote{\url{http://deep.ps.uci.edu/spec2d/}} used by the DEEP2 survey project, 
which measures redshifts using cross-correlation with templates that contain absorption and emission line spectra. Then 3 persons check the quality of the spec-$z$ fitting result individually. When the fitted spec-$z$ is not matched by the location of spectral lines, we re-fit the spec-$z$ to match the spectral lines within a smaller wavelength range. We assign the quality flag ($qz$) of the spec-$z$ with the same criterion as the DEEP2 survey \citep{Willmer2006, 2013ApJS..208....5N}. Finally, we check the spec-$z$ catalogue from the three persons and find good consistency. For the targets with inconsistent spec-$z$ results, their reliability will be changed from the average $qz$ estimated from the visual inspection to $qz-1$. Low-$z$ galaxies have many observable emission and absorption line features in the observation window, thus the quality would be either identified by two clear lines, or no available feature to help the identification. So almost all galaxies in our sample have the quality flag 4, or unreliable quality flag as 1 or 2.

We obtained in total 6000 spectra, 2753 of which have high reliable quality ($qz \geq3$). The upper panel of Fig. \ref{specz_hist} shows the observed color magnitude diagram and the histogram of our full sample observed with MMT/Hectospec (in black color) and the targets with $qz\geq3$ (in red color). The redshift identification success rate is the ratio between the red and black histogram respectively, which is about 50\% for our sample with r band 22 magnitude, or u-r $\gtrsim 2.5$. Red targets usually have weak emission lines thus the more difficult to identify the redshift. Galaxies with low surface brightness in the fiber aperture are also challenging to obtain spectra with S/N high enough to identify the redshifts. The success rate is also related to the exposure time, weather, moon phase, etc.

\startlongtable
\begin{deluxetable*}{lccccc}
\tabletypesize{\footnotesize}
\tablewidth{0pt}
\tablecaption{Redshift Catalogue of galaxies we observed in the XMM-LSS, ELAIS-N1 and DEEP2-3 fields.\label{tab_label}}
\tablecomments{Table 1 is published in its entirety in the machine-readable format.
A portion is shown here for guidance regarding its form and content.}
\tablehead{ID  & RA(J2000) & Dec(J2000) &  $r_{\rm aper}$(1") &  $z_{\rm spec}$ &  $qz$}
\startdata
J022455.7-045217.2  &  36.23213053  &  -4.87143230 & 20.05  &   0.435135  $\pm$ 0.000083 &     4 \\
J022404.9-044628.7  &  36.02022171  &  -4.77463340 & 20.62  &   0.069242  $\pm$ 0.000011 &     4 \\
J022422.8-050130.2  &  36.09491587  &  -5.02506590 & 21.18  &   0.361262  $\pm$ 0.000150 &     3 \\
J022415.1-044629.2  &  36.06300831  &  -4.77476690 & 20.39  &   0.313309  $\pm$ 0.000019 &     4 \\
J022451.1-045053.0  &  36.21307254  &  -4.84805010 & 20.30  &   0.151206  $\pm$ 0.000017 &     4 \\
...\\
\enddata
\end{deluxetable*}

The lower panel of Fig. \ref{specz_hist} shows the redshift distribution of galaxies with $qz \geq 3$. (164 SWIRE targets have reliable spec-$z$ at the range of $z_{\rm spec}<0.6$ in our survey. We do not treat the SWIRE targets separately in the following analyses.) Our targets are selected from $r < 22$, which is highly complete because the optical survey depth is about $r$ band 25 – 27 mag, much deeper than $r = 22$. The observed-frame $u-g$, $r-i$ color-color selection aims to select the galaxies at $z_{\rm spec}<0.35$, however, there is no clear division between the $z_{\rm spec} > 0.35$ and $z_{\rm spec} < 0.35$ galaxies in the color-color diagram, thus there would be some low-redshift galaxies outside the selection box, or vice versa. Our results show that about 20\% (553 out of 2753) of the galaxies in our selection box are at $z_{\rm spec} > 0.35$, and if the redshift distribution of the targets close to the color-color lines is symmetric, we can expect an 80\% completeness for the broad band color-color selection method. The redshift identification success rate also limits our sample completeness especially for the faint or red galaxies. Based on the histograms in Fig. \ref{specz_hist}, targets selected by the color-color diagram have a success rate about 80\% for the galaxies $r<21$ or observed-frame color $u-r<2$, and only 50\% for the galaxies about $21 < r < 22$ and observed-frame $u-r > 2.5$. The redshift distribution histogram also shows the presence of large scale clustering structures at several redshift bins, which is also revealed in the subplot of Fig. \ref{ugri_selection}. The catalogue is listed in Table \ref{tab_label}, and can be accessed through the webpage of this survey\footnote{\url{http://mips.as.arizona.edu/~cnaw/Faint_Low_z/}}.

We show all of our spectra with $qz \geq 3$ in Fig. \ref{spec-image} and denote the position of prominent lines. In addition to the emission lines, several prominent absorption lines can be identified such as the H$\delta$, K, H, G-band, Mgb and NaD. These lines are important in revealing the post-starburst features, age, outflows and metallicity. We fit the emission lines by one Gaussian profile plus n-order polynomial to model the continuum.. We also measure the $D_n4000$ parameter \citep{Balogh1999}. To reduce the uncertainty of the flux calibration, we estimate the metallicity by the [NII]/H$\alpha$ and the Baldwin, Phillips 
\& Telervich (BPT) diagram \citep{1981PASP...93....5B} of [NII]/H$\alpha$ v.s. [OIII]/H$\beta$ so that the flux ratio always comes from two emission lines at close wavelengths.

\subsection{spectrum examples}

We show some examples of our MMT/Hectospec spectra in Fig. \ref{spec-example}. The $r$ band magnitude in each panel increases from the top to the bottom panel. For 1.5 hours exposure time, S/N of the MMT/Hectospec spectra are high enough to identify the emission and absorption lines for the galaxies brighter than $r = 21$ AB mag, and bright enough to identify emission lines for $21<r<22$. For the faint emission-line galaxies, we can only identify the redshifts and cannot have reliable D$_n$4000 measurements. The flux calibration from F-star can normalize the continuum flux, but do not calibrate the shape of the continuum very well. The data from the two grey nights in 2017B are affected by moonlight, so we can only get reliable redshift.

\begin{figure}[ht!]
\centering
\includegraphics[width=0.47\textwidth]{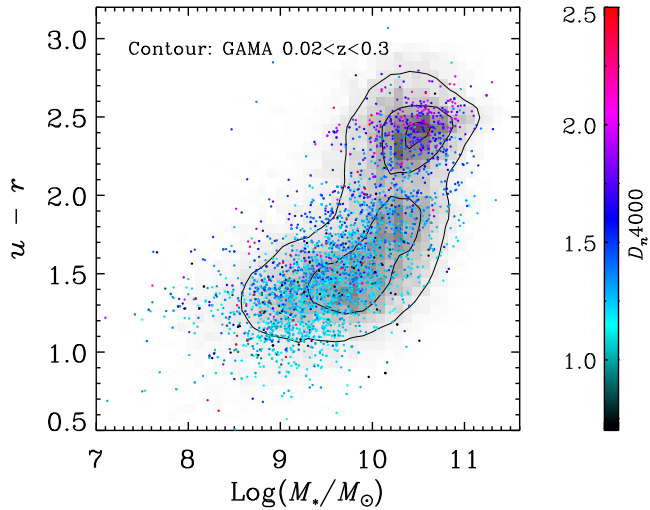}
\caption{colour-magnitude and colour stellar mass plots. We compare with the GAMA sample at $0.02<z_{\rm spec}<0.3$. We colour our results with the $D_n4000$ value. 
}\label{colour-mass}
\end{figure}

\subsection{Multi-wavelength SED fitting}

The survey fields have been covered by CLAUDS in u band, HSC in $g, r, i, z, y$ band and also covered by the near Infrared (NIR) images from the VISTA Deep Extragalactic Observations (VIDEO) survey \citep{2013MNRAS.428.1281J}  in the XMMLSS field, the UKIRT Infrared Deep Sky Survey / Deep Extragalactic Survey \citep[UKIDSS/DXS,][]{2007MNRAS.379.1599L} in the ELAIS-N1 field and the VISTA-CFHT Stripe 82 Survey  \citep[VICS82,][]{2017ApJS..231....7G} in the DEEP2-3 field. We match our redshift catalogue to the CLAUDS and HSC dr2 catalogue, which have much deeper limiting magnitude ($\sim$ 27), and more reliable photometry. Some of our targets have bright neighbouring stars in the HSC catalogue, which were flagged or not contained in the HSC catalogues. For these objects, we use the photometry from the target selection catalogues, which have brighter saturation magnitudes than the HSC survey, and are much less affected from the bright stars. We also match our catalogue to the archive UKIRT $J$ and $Ks$ bands catalogues to have better coverage in the NIR bands in the final multi-wavelength catalogue. 

\begin{figure}
\centering
\includegraphics[width=0.47\textwidth]{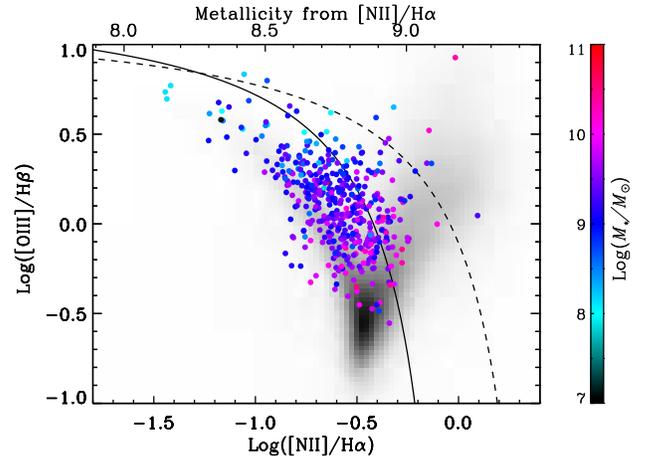}
\caption{BPT diagram of our sample with emission lines. The greyscale shows the distribution of SDSS galaxies. We also plot the Kauffmann and Kewley lines \citep{2001ApJ...556..121K, 2003MNRAS.346.1055K}, which classify galaxies as being mainly ionized by active galactic nucleus or star formation. The upper horizontal axis show the metallicity that derived from the [NII]/H$\alpha$ \citep{Curti2020}.
}\label{BPT}
\end{figure}

\begin{figure}
    \centering
    \includegraphics[width = 0.47\textwidth]{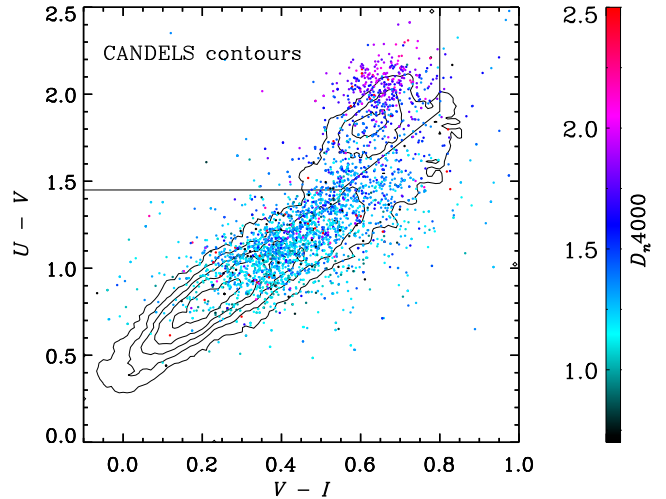}
    \caption{The $U-V$ v.s. $V-I$ diagram of our sample overploted on the contour of CANDELS UVI results. The solid line separates the quiescent galaxy population (upper left) from the dusty (upper right) and star forming (lower left) galaxy populations. We use the galaxy sample from the CANDELS catalog with $z_{\rm phot}< 1$. }
    \label{UVI}
\end{figure}

We derive the absolute magnitude by the definition of $m_R = M_Q +DM + K_{QR}$ \citep{2002astro.ph.10394H}, where the $m_R$ is the observed magnitude in band $R$, $M_Q$ is the absolute magnitude in band $Q$ in the emitted frame, $DM$ is the distance modulus and the $K_{QR}$ is the $K$ correction factor between the observed $R$ band and the rest-frame $Q$ band. The $K$-correction factor of the absolute magnitude of our sample is derived by the \citet{1980ApJS...43..393C} and \citet{1996ApJ...467...38K} template spectra with the $K$-correction formulae given by \citet{2002astro.ph.10394H}. We show the results in Fig. \ref{abs_mag} and compare with the previous low-redshift spec-$z$ survey results from SDSS and GAMA. We select 20000 SDSS sources with random RA, Dec and the full GAMA catalogue \citep{2011MNRAS.413..971D, 2016MNRAS.455.3911D} in the G09, G12, G15 fields as a comparison sample. Within the same redshift range, our results reach about 2 magnitudes fainter than the previous works. Galaxies in our sample with $M_r\simeq -18$ are located at $z_{\rm spec} \sim 0.1$, a distance where peculiar motions can be neglected. Nevertheless, the SDSS and GAMA surveys mainly sample galaxies with $M_r\simeq -18$  at redshifts $z < 0.05$. Therefore, our sample reaches lower stellar masses than the previous surveys at higher redshift. 

We fit the SED by {\tt fastpp}\footnote{https://github.com/cschreib/fastpp}\citep{2009ApJ...700..221K} using the broad-band photometry and our spec-$z$s to measure the stellar mass. We adopt the \citet{2003MNRAS.344.1000B} stellar population synthesis models, Chabrier initial mass function \citep{2003PASP..115..763C}, and the \citet{2000ApJ...533..682C} dust extinction law with the attenuation in the range of $ 0<A_{\rm v}<3$. Fig. \ref{colour-mass} shows the absolute $u-r$ colour and stellar mass diagram. We also show the contours of the colour-mass diagram of GAMA of the galaxies at $0.02<z_{\rm spec}<0.3$. Our results are consistent with GAMA, while the stellar masses are a bit lower than GAMA.

\section{Results}
\subsection{Emission line diagnosis}
The BPT diagram is widely used to identify the emission line ionisation source based on the line ratios \citep{1981PASP...93....5B, 2006MNRAS.372..961K}. We show the BPT diagram in Fig. \ref{BPT} where the colours represent different stellar mass ranges. We also show the result from SDSS to show the typical values of the low-$z$ galaxies. Nearly all the galaxies in our sample are star-forming galaxies. Gas-phase metallicities can be derived from the strong line ratios \citep{2008ApJ...681.1183K, Maiolino2019}. We denote the metallicity that is derived from [NII]/H$\alpha$ \citep{Curti2020} in the upper x-axis. 
The mass-metallicity trend \citep{2004ApJ...613..898T} can also be seen in our sample, where the low metallicity targets have stellar masses of $\sim 10^8M_\odot$.

\begin{figure}
    \centering
    \includegraphics[width=0.47\textwidth]{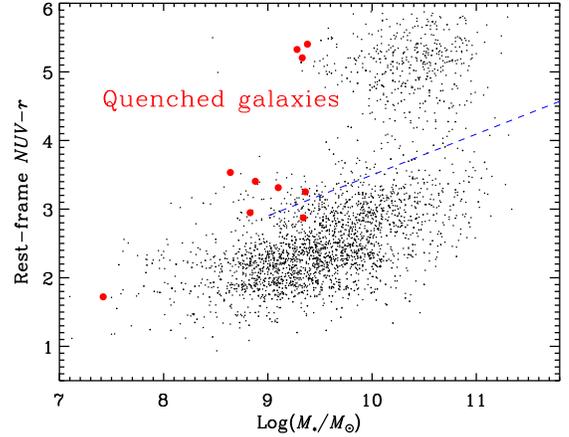}
    \caption{$NUV-r$ v.s stellar mass diagram of our sample. The blue dashed line represents the criterion of \citet{2016ApJ...819...91P} to separate red from star-forming galaxies. This line is generated from the SDSS data with stellar mass lower to $10^9M_\odot$. The red filled circles are the low mass galaxies ($M_*<10^{9.5}M_\odot$) with no clear emission line, which are the quenched galaxies we found in this spec-$z$ survey.}
    \label{NUV-r-mass}
\end{figure}

\subsection{rest-frame U - V v.s. V - I diagram}

The UVJ diagram (rest-frame U - V v.s. rest-frame V - J) has been widely applied to help constrain galaxy properties because extinction and age vary in orthogonal directions \citep{2005ApJ...624L..81L, 2010ApJ...713..738W, Fang2018, 2019ApJ...880L...9L}. However, because of the limited NIR coverage of our sample, we followed \citet{Wang2017, Liu2018} by using the UVI distribution (rest-frame U - V v.s. V - I, Fig. \ref{UVI}) to separate dusty from old populations. We define the old population region based on the quiescent galaxies selected from the UVJ diagram. Our result shows a similar distribution as the the $z_{\rm phot}<1$ sample from Cosmic Assembly Near-infrared Deep Extragalactic Legacy Survey \citep[CANDELS,][]{2011ApJS..197...36K, 2011ApJS..197...35G}, which are denoted by the contours.

\begin{figure}
    \centering
    \includegraphics[width=0.47\textwidth]{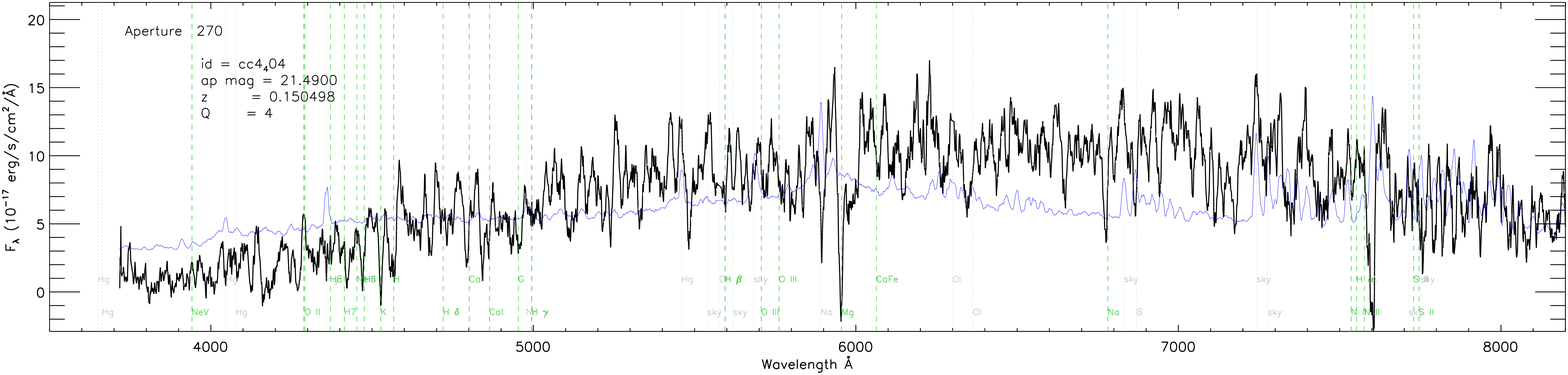}
    \includegraphics[width=0.47\textwidth]{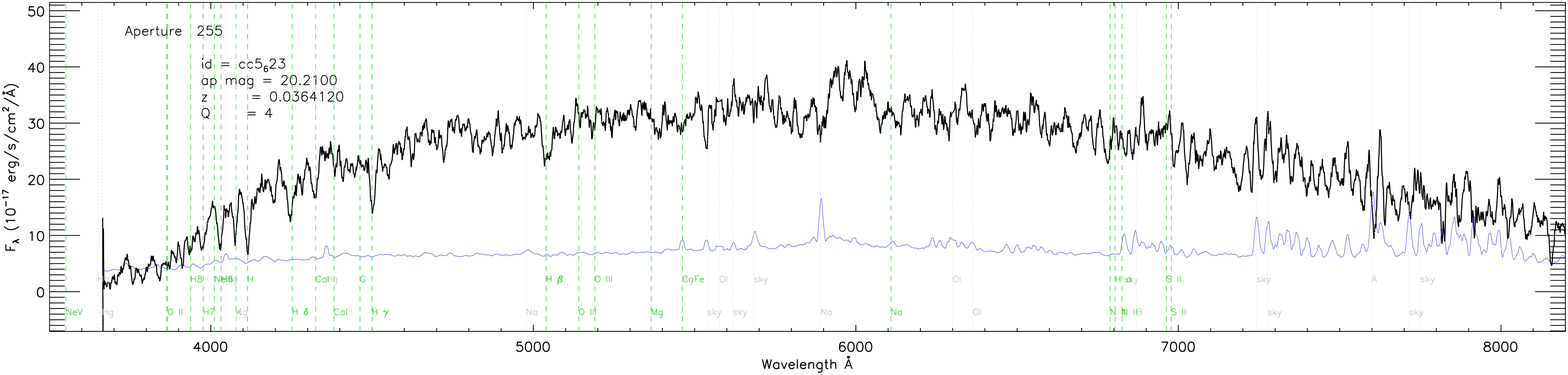}
    \includegraphics[width=0.47\textwidth]{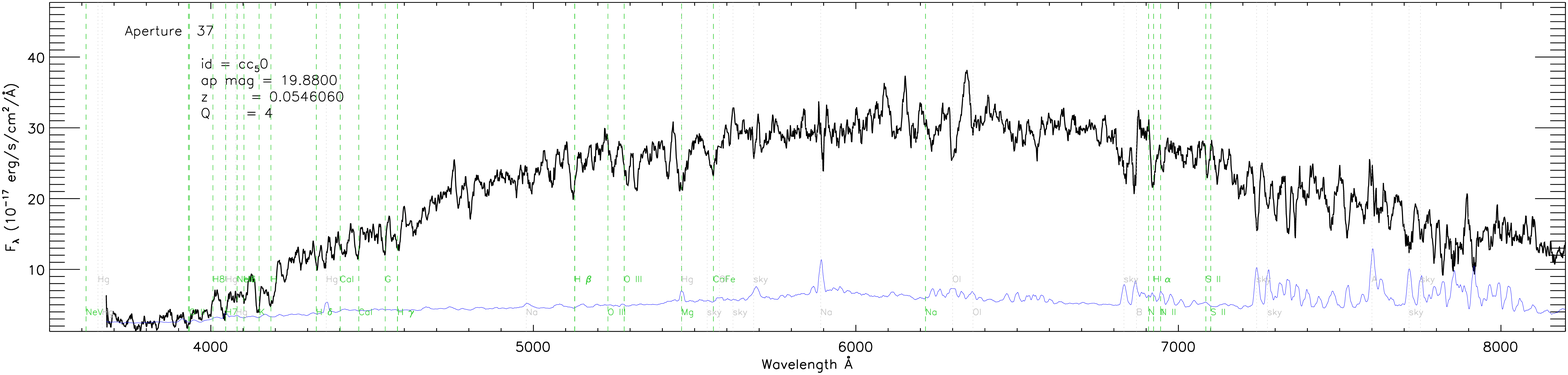}
    \includegraphics[width=0.47\textwidth]{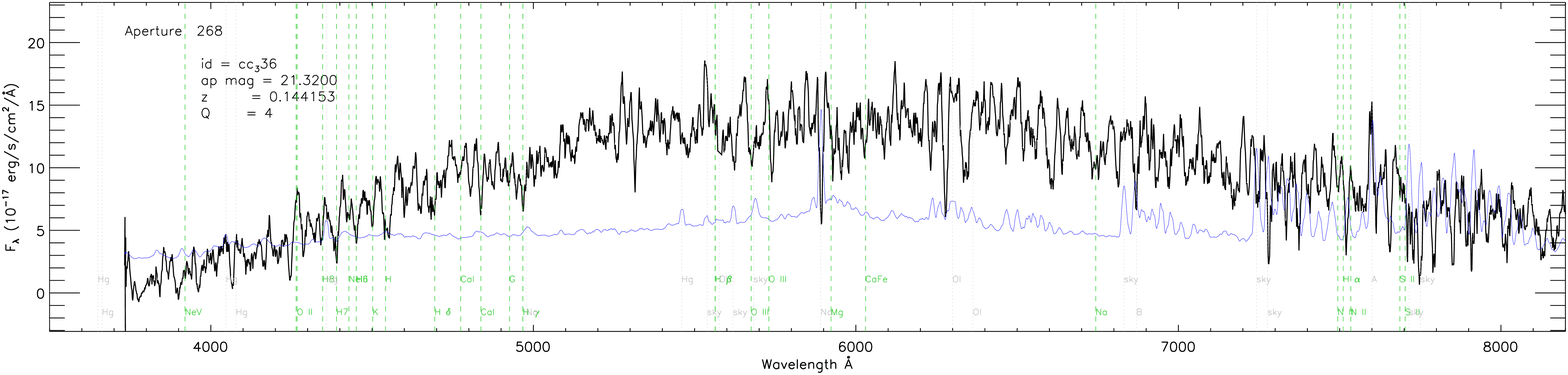}
    \caption{Example of spectra of low mass galaxies ($M_*<10^{9}M_\odot$) with weak emission lines. }
    \label{lmass8}
\end{figure}

\begin{figure}
    \centering
    \includegraphics[width=0.47\textwidth]{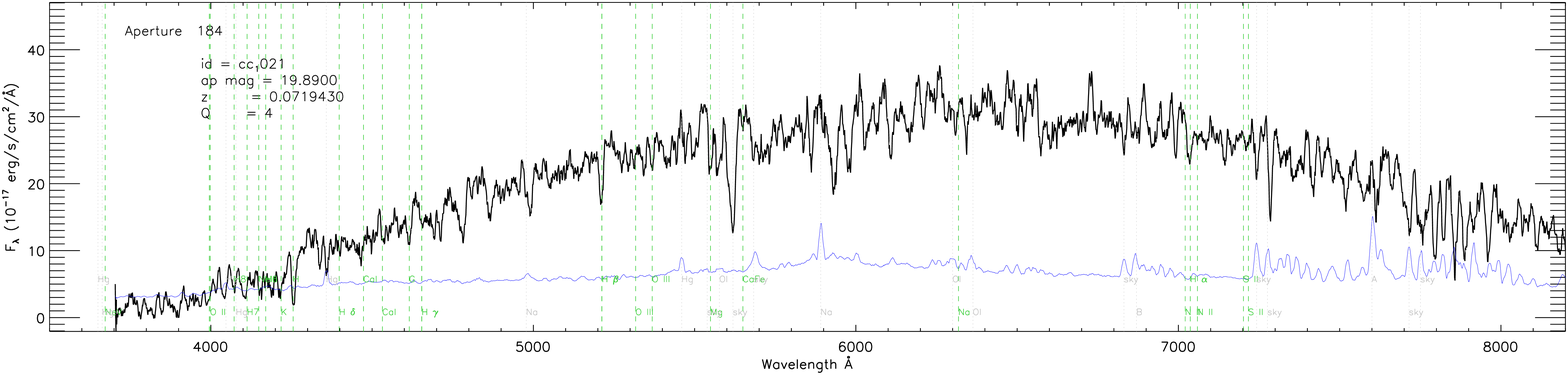}
    \includegraphics[width=0.47\textwidth]{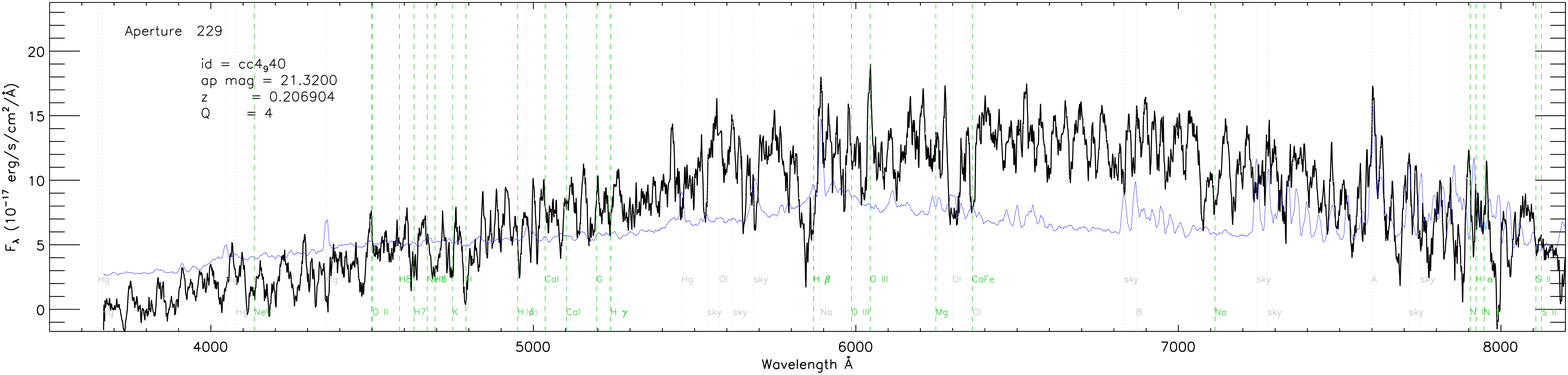}
    \includegraphics[width=0.47\textwidth]{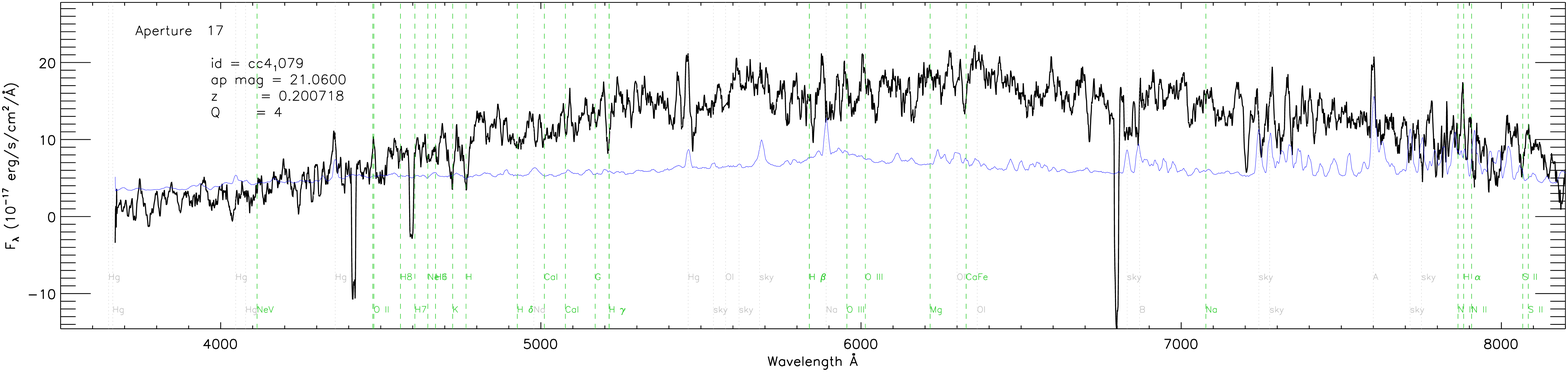}
    \includegraphics[width=0.47\textwidth]{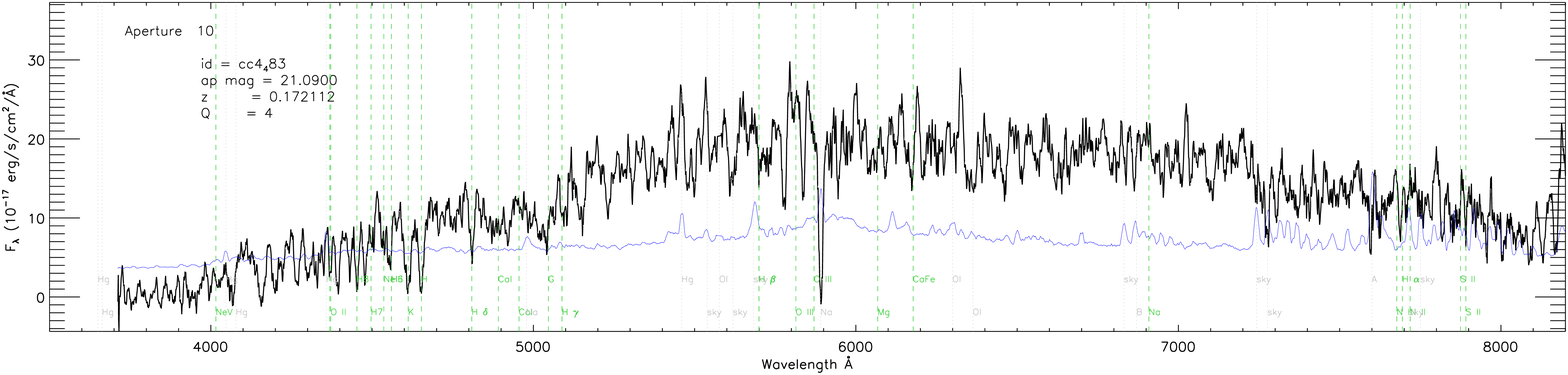}
    \includegraphics[width=0.47\textwidth]{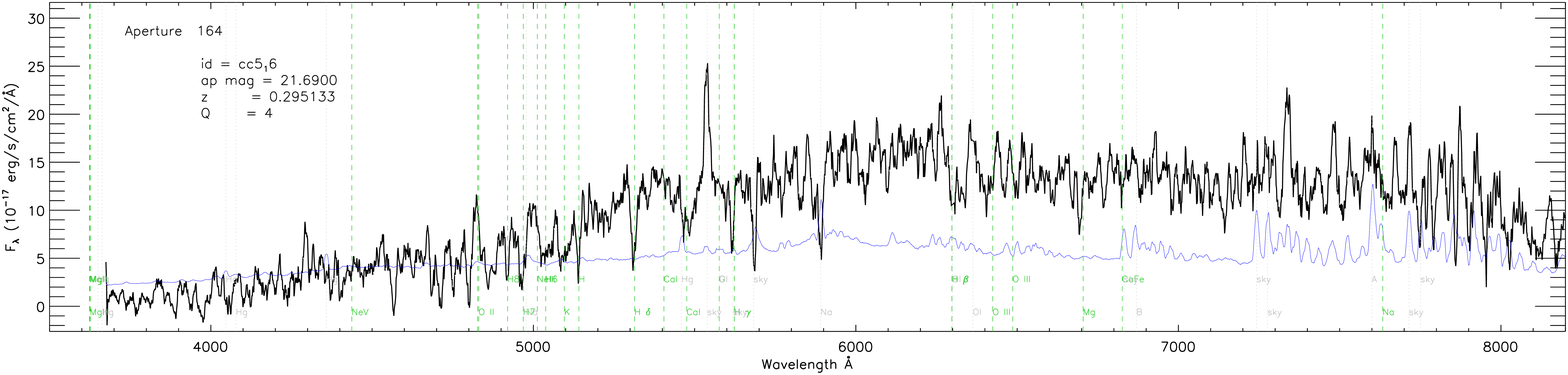}
    \includegraphics[width=0.47\textwidth]{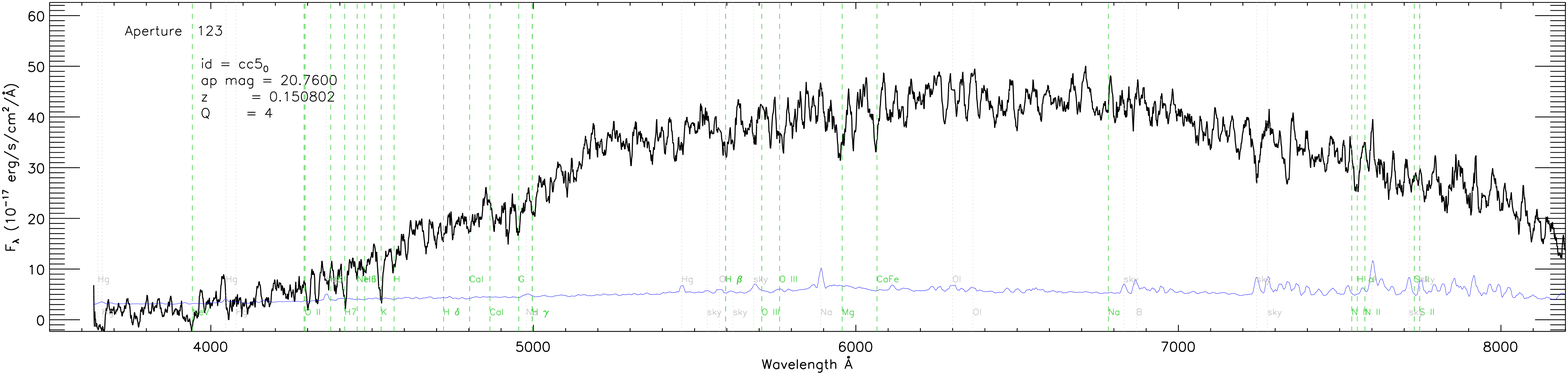}
    \caption{Example of spectra of low mass galaxies ($10^{9}M_\odot<M_*<10^{9.5}M_\odot$) with weak emission lines. }
    \label{lmass9}
\end{figure}

\subsection{Absorption line galaxies: quenching of the low mass galaxies}
Quenching of low mass galaxies is commonly studied in galaxy clusters\citep{2014ApJ...783..120T, 2017ApJ...836..120R, 2018ApJ...856...69D}, where the massive galaxies reduce the gas supply to dwarf galaxies by ram-pressure, tidal interactions or other mechanisms \citep{2019ARA&A..57..375S}. Fig. \ref{BPT} shows that most of our galaxies are star-forming galaxies at $\sim 10^9M_*/M_\odot$. So we can use our sample to understand the quenching features of field galaxies at $\sim 10^9M_*/M_\odot$.

On the other hand, previous studies show that the fraction of isolated quenching galaxies in the field is extremely small at the low mass end ($10^7<M_*/M_\odot<10^9$, or, $-18<M_r<-12$) \citep{Geha2012}. In this case, the low mass galaxies should either sustain the current star formation, grow into massive galaxies, or be merged into nearby massive galaxies. However, star formation history analyses of the ultra-faint local dwarf galaxies \citep{Brown2014} show that 80\% stars were formed at about $z\sim 6$ and 100\% stars were formed at $z \sim 2$. Therefore, we could expect to find the low mass quenched galaxies at low-redshift.

The quenching and star-forming galaxy population can be shown clearly in the rest-frame $NUV-r$ v.s. stellar mass diagram \citep{2005ApJ...619L..39S, 2014SerAJ.189....1S, 2016ApJ...819...91P}, where the rest frame $NUV$ flux is proportional to the star formation rate, therefore the $NUV-r$ roughly represents the specific star formation rate. One of the advantages of the $NUV-r$ diagram is the clear separation between the galaxy populations\citep{2014SerAJ.189....1S}. Fig. \ref{NUV-r-mass} shows the  $NUV-r$ v.s. mass of our sample, where we can see a clear separation between the blue and red population, and the green valley where the galaxies might be post-starburst galaxies. 

We searched for the presence of strong Balmer absorption and weak or no emission line flux in the spectra of all galaxies with $\log(M_*/M_\odot)<9.5$. We find 10 galaxies with no clear emission lines (S/N $<$ 3) and these are shown in Fig. \ref{lmass8} for $\log(M_*/M_\odot)<9$, Fig. \ref{lmass9} for  $9<\log(M_*/M_\odot)<9.5$, and in Fig. \ref{NUV-r-mass} with red dots. The very weak emission lines, the strong H$\delta$ (if detected) and the H, K lines of the spectrum indicate a lack of ionized gas, quenched from starburst phase and may evolve into the red population. We show the star-forming galaxy separation criteria from \citet{2016ApJ...819...91P}, where they only consider the galaxy sample with a stellar mass larger than $10^9M_\odot$ as a reference. The $NUV-r$ colour of these galaxies shows that the 10 weak emission-line galaxies are above the main sequence of the star-forming galaxies in Fig. \ref{NUV-r-mass}, more likely in the green valley. We observed the absorption line galaxy with the lowest stellar mass in Fig. \ref{NUV-r-mass} by using the Hale telescope/DBSP  (TAP2017B, ID: 01. PI: Cheng Cheng) confirming the 0.036 spectroscopic redshift of this galaxy and find no evidence of emission lines. We also confirme that there is no nearby galaxy within 50 kpc brighter than this target. \citet{Geha2012} have shown that the isolated low mass quenched galaxies are rare. A detailed study of these low mass absorption line galaxies, such as whether they are isolated galaxies, will be presented in a forthcoming paper.

\begin{figure}
    \centering
    \includegraphics[width = 0.47\textwidth]{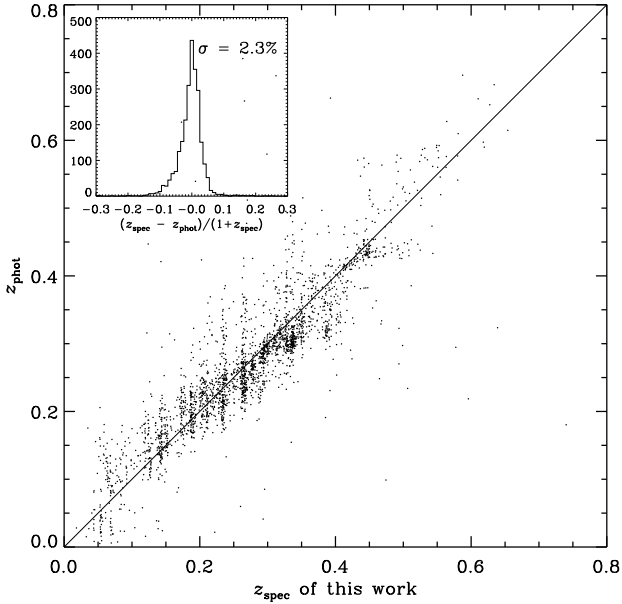}
    \caption{Our $z_{\rm spec}$ v.s. the $z_{\rm phot}$ derived from the CLAUDS + HSC dr2 SEDs (Desprez et al. in prep). The $z_{\rm phot}$ has a scatter about 2.3\% and offset about 0.006. The large scale clustering structures shown in $z_{\rm spec}$ are not recovered in the $z_{\rm phot}$ results.
    }
    \label{photz}
\end{figure}

We also take advantage of our accurate phot-$z$ to study the low mass quenching population. The ongoing and upcoming multi-wavelength deep survey projects such as HSC and Rubin Observatory's Legacy Survey of Space and Time \citep[LSST,][]{2009arXiv0912.0201L} data will produce us a vast sample of high-quality photometry catalogues. Meanwhile, phot-$z$ techniques such as machine learning and empirical template fitting, as well as the hybridization of several methods \citep{2019NatAs...3..212S} can provide accuracies of 2\% or even better \citep{2018PASJ...70S...9T, 2019MNRAS.488.4565Z, 2020arXiv200301511N, 2020arXiv200103621S}, which can be used to enlarge the sample of low-redshift low mass passive galaxies for follow up spectroscopic identification. We compare the phot-$z$ results of our sample from the CLAUDS + HSC dr2 catalog (Desprez et al. in prep) and our spec-$z$ results in Fig. \ref{photz}. The phot-$z$ of our sample can recover the spec-$z$ very well with a scatter $\sigma_{\rm NMAD}$ about 2.3\% and a redshift offset about 0.006. The large scale clustering structures shown in $z_{\rm spec}$ (also see Fig. \ref{specz_hist} right panel) are not recovered in the phot-$z$ results, implying the limitation of studying the cosmology large scale structure with phot-$z$ data \citep[also see][]{2021ApJ...909..129S}. In the forthcoming work, we will study the low mass red galaxies based on the spec-$z$ and the phot-$z$ to explore the quenching process in the low mass galaxies.

\section{Summary}

We select a sample of $18< r < 22$ low-redshift field galaxies ($0.03 \lesssim z_{\rm spec}\lesssim0.5$) and carry out a spectroscopic redshift survey using MMT/Hectospec. The redshift distribution validates our target selection method, and we acquired spectra for 2753 galaxies. In this paper, we release the spec-$z$ catalogue including ra, dec, spec-$z$ and $qz$. We demonstrate that our selection method is effective to identify low-mass quenched galaxies. 
The Spec-$z$ catalogue was used to calibrate the phot-$z$s in three surveyed fields expanding the sample of low-mass galaxies. 
These combined (spectrocopic and photometric redshift) catalogues are being used to analyse the population of low-mass field galaxies, which are the subject of a forthcoming paper.

\acknowledgments
We thank the referee for careful reading and constructive suggestions. C.C. would like to thank Daniel Fabricant for approving the usage of MMT/Hectospec. We thank  Zheng Cai, Song Huang and Cheng Li for helpful discussion. This work is supported by the National Key R\&D Program of China grant 2017YFA0402704 and by the National Natural Science Foundation of China, No. 11803044, 11933003. This work is sponsored (in part) by the Chinese Academy of Sciences (CAS), through a grant to the CAS South America Center for Astronomy (CASSACA). We acknowledge the science research grants from the China Manned Space Project with NO. CMS-CSST-2021-A05. This research uses data obtained through the Telescope Access Program (TAP). Observations reported here were obtained at the MMT Observatory, a joint facility of the University of Arizona and the Smithsonian Institution. 

These data were obtained and processed as part of the CFHT Large Area U-band Deep Survey (CLAUDS), which is a collaboration between astronomers from Canada, France, and China described in Sawicki et al. (2019, [MNRAS 489, 5202]).  CLAUDS is based on observations obtained with MegaPrime/ MegaCam, a joint project of CFHT and CEA/DAPNIA, at the CFHT which is operated by the National Research Council (NRC) of Canada, the Institut National des Science de l'Univers of the Centre National de la Recherche Scientifique (CNRS) of France, and the University of Hawaii. CLAUDS uses data obtained in part through the Telescope Access Program (TAP), which has been funded by the National Astronomical Observatories, Chinese Academy of Sciences, and the Special Fund for Astronomy from the Ministry of Finance of China. CLAUDS uses data products from TERAPIX and the Canadian Astronomy Data Centre (CADC) and was carried out using resources from Compute Canada and Canadian Advanced Network For Astrophysical Research (CANFAR).

This work is also based in part on data collected at the Subaru
Telescope and retrieved from the HSC data archive system, which
is operated by the Subaru Telescope and Astronomy Data Center at
National Astronomical Observatory of Japan. The Hyper Suprime-Cam (HSC) 
collaboration includes the astronomical communities of Japan and 
Taiwan, and Princeton University. The HSC instrumentation and 
software were developed by the National Astronomical Observatory 
of Japan (NAOJ), the Kavli Institute for the Physics and Mathematics 
of the Universe (Kavli IPMU), the University of Tokyo, 
the High Energy Accelerator Research Organization (KEK), the Academia 
Sinica Institute for Astronomy and Astrophysics in Taiwan (ASIAA), 
and Princeton University. Funding was contributed by the FIRST 
programme from Japanese Cabinet Office,
the Ministry of Education, Culture, Sports, Science and Technology
(MEXT), the Japan Society for the Promotion of Science (JSPS),
Japan Science and Technology Agency (JST), the Toray Science
Foundation, NAOJ, Kavli IPMU, KEK, ASIAA, and Princeton
University. 

Based on observations obtained with MegaPrime/MegaCam, a joint project of CFHT and CEA/IRFU, at the Canada-France-Hawaii Telescope (CFHT) which is operated by the National Research Council (NRC) of Canada, the Institut National des Science de l'Univers of the Centre National de la Recherche Scientifique (CNRS) of France, and the University of Hawaii. This work is based in part on data products produced at Terapix available at the Canadian Astronomy Data Centre as part of the Canada-France-Hawaii Telescope Legacy Survey, a collaborative project of NRC and CNRS.

\bibliography{aastex63}{}
\bibliographystyle{aasjournal}

\end{document}